\documentclass[aps,prb,twocolumn,superscriptaddress]{revtex4-2}
\usepackage{xcolor}
\usepackage{qcircuit,amsmath}
\usepackage{grffile}
\usepackage[export]{adjustbox}
\usepackage{lineno}
\usepackage{xcolor}
\usepackage{longtable}
\usepackage{braket}
\usepackage{array}
\usepackage{multirow}
\usepackage{enumerate}
\usepackage{amsmath,empheq}
\usepackage{amssymb}
\usepackage{amsthm}
\usepackage{times}
\usepackage{multirow}
\usepackage{MnSymbol}

\definecolor{LinkColor}{RGB}{77,77,192}
\usepackage[colorlinks, citecolor=blue]{hyperref}
\hypersetup{linkcolor=LinkColor,urlcolor=LinkColor,citecolor=LinkColor,pdfstartview={FitH},urlcolor=LinkColor}

\usepackage[utf8]{inputenc}
\usepackage{color}
\usepackage{tabularx}
\usepackage{algorithm}
\usepackage{algpseudocode}
\usepackage{graphbox}
\usepackage{amsfonts}
\usepackage{dcolumn}
\usepackage{bm}
\usepackage{epstopdf}
\usepackage{enumitem}

\begin{document}

\title{Quantum Mpemba effect in long-ranged U(1)-symmetric random circuits} 

\author{Han-Ze Li}
\affiliation{Institute for Quantum Science and Technology, Shanghai University, Shanghai 200444, China}
\affiliation{Department of Physics, National University of Singapore, Singapore 117542}

\author{Ching Hua Lee}
\affiliation{Department of Physics, National University of Singapore, Singapore 117542}

\author{Shuo Liu}
\email{sl6097@princeton.edu}
\affiliation{Department of Physics, Princeton University, Princeton, New Jersey 08544, USA}
\affiliation{Institute for Advanced Study, Tsinghua University, Beijing 100084, China} 

\author{Shi-Xin Zhang}
\email{shixinzhang@iphy.ac.cn}
\affiliation{ Institute of Physics, Chinese Academy of Sciences, Beijing 100190, China}

\author{Jian-Xin Zhong}
\email{jxzhong@shu.edu.cn}
\affiliation{Institute for Quantum Science and Technology, Shanghai University, Shanghai 200444, China}
% \date{\today}

\begin{abstract}
The Mpemba effect, where a state prepared farther from equilibrium relaxes faster to equilibrium than one prepared closer, has a quantum counterpart where relaxation is resolved by conserved charge. However, the fate of the quantum Mpemba effect in systems with long-range interactions remains an open question. Here, we study the quantum Mpemba effect in long-ranged, U(1)-symmetric random unitary circuits. Using annealed R\'enyi-2 entanglement asymmetry computed via replica tensor networks and exact diagonalization, we track the symmetry restoration from three types of tilted product states: ferromagnetic, antiferromagnetic, and ferromagnetic with a central domain wall. The quantum Mpemba effect is present for tilted ferromagnetic states at all interaction ranges, but absent for tilted antiferromagnetic states, and occurs for the domain-wall state only in effectively short-ranged circuits, where the Mpemba time $t_{\rm M}$ is found to scale with the subsystem size $N_A$ as $t_{\rm M}\!\sim\!N_{A}^{\,z}$, with the dynamical exponent $z=\min(\alpha-1,2)$. 
These results reveal how the quantum Mpemba effect is governed by the interplay between interaction range and initial-state charge bias in long-ranged chaotic systems.

\end{abstract}

\maketitle

\section{Introduction}
Physical systems driven out of equilibrium tend to relax towards a steady state determined by the environment~\cite{DEMIREL2019573}.
Conventional wisdom holds that a hotter object cools more slowly because it must pass through every intermediate temperature.
However, under certain conditions, a counterintuitive phenomenon emerges: a system initially farther from equilibrium can approach equilibrium faster than one initially closer to it. 
This atypical relaxation phenomenon, known as the Mpemba effect~\cite{MpembaOsborne1969}, has persistently challenged classical thermodynamic intuition since Mpemba's modern account of it in the 1960s, sparking decades of debate over its microscopic origin~\cite{C4CP03669G,PhysRevLett.119.148001,PhysRevE.106.034131,Ghosh2025,PhysRevLett.134.107101}.

Over the past decade, the scope of the Mpemba effect has expanded well beyond classical heat-transfer scenarios. In classical systems, it has been observed across a wide range of
platforms~\cite{Auerbach1995Mpemba,Esposito_2008,a1,a2,a3,a4,a5,a6,a7,a8,a9,a11,a12,a13,a14,a15,a16,hou1,hou2,Longhi1,Longhi2,Longhi3}, unified by a simple modal picture: relaxation can be decomposed into modes and equilibration accelerates when the initial state suppresses the slowest one, typically occurring for hotter starts. 
However, the classical Mpemba effect remains under active debate~\cite{Burridge2016,Henry}, owing to the complexity of the underlying dynamics, which also raises the question of whether analogous phenomena can emerge in quantum systems.

In the quantum regime, the Mpemba effect emerges in both open~\cite{PhysRevLett.131.080402,PhysRevLett.133.010403,1,2,3,4,6,7,9,16,17,add1,add2,add3,add4,add5,add_inverseME,add6,add7,add8,add9,add10,add11,add12,add13,longhi4,oct2,oct3,zhang2025quantummpembaeffectinduced} and isolated~\cite{Ares2023first,0,5,8,10,12,13,14,15,xu2025observationmodulationquantummpemba,18,19,21,22,liu1,liu2,xhek,add_inter1,add_inter2,add_inter4,add_inter5,oct1,oct4} systems, exhibiting a rich spectrum of relaxation behaviors, including the inverse~\cite{PhysRevLett.133.010403,12,add_inverseME} and strong~\cite{Zhang_2025} Mpemba effects.
In the quantum version, 
% it is emphasizing that
the accelerated relaxation of the Mpemba effect is governed by interparticle correlations, quantum fluctuations, and the choice of initial conditions. Key frontiers are to pinpoint when and why the anomalous acceleration occurs, and how it can be controlled.
Among these quantum realizations, we focus on the type of quantum Mpemba effect (QME) in isolated quantum systems from the perspective of symmetry restoration~\cite{Ares2023first}. Hereafter, the abbreviation ``QME'' will specifically refer to this type of quantum Mpemba effect.
It depicts the counterintuitive phenomenon in the context of symmetry restoration~\cite{SR9,SR8,SR7,SR6,SR5,SR4,SR3,SR2,SR1} in closed quantum systems: a more asymmetric initial state can restore the subsystem symmetry faster than one that is less asymmetric (see recent reviews~\cite{Ares_2025,11}).
Under a global quench of a U(1)-symmetric Hamiltonian, strongly symmetry-broken initial states exhibit accelerated symmetry restoration, a hallmark of QME that has been studied for its connection to integrability in one-dimensional spin chains. Moreover, QME has been demonstrated across a broad range of platforms, from integrable~\cite{add_inter1,add_inter2,SR3,add_inter4,add_inter5,PhysRevLett.133.010401} to chaotic models~\cite{liu1,xhek,Bertini,8,SR9}, and in other setups~\cite{Medina_2025,add_quantum_battery,24,25,26}.
Experimentally, noise- and disorder-robust QME has also been observed on superconducting~\cite{xu2025observationmodulationquantummpemba} and trapped-ion platforms~\cite{PhysRevLett.133.010402}.

Despite its substantial prominence, a question remains only partially answered:
How does the QME evolve in a rapidly thermalizing long-ranged interacting chaotic system? Although the QME has been observed in some long-range setups recently~\cite{xu2025observationmodulationquantummpemba,14,Matthew,wang2024mpembameetsquantumchaos,22}, its quantitative dependence on the choice of initial-state and on the interaction range remains largely unexplored. Given that many experimental platforms, including trapped-ion simulators~\cite{PhysRevLett.133.010402} and Rydberg-atom arrays~\cite{lukin1,lukin2,lukin3,lukin4,lukin5,lukin6,PhysRevLett.131.080403}, naturally feature long-range interactions~\cite{RevModPhys.95.035002}, understanding the general framework to characterize QME in such long-range systems is
both of theoretical interest and experimental relevance.
% timely and essential.

\begin{figure}[bt]
\hspace*{-0.5\textwidth}
\centering
\includegraphics[width=0.5\textwidth]{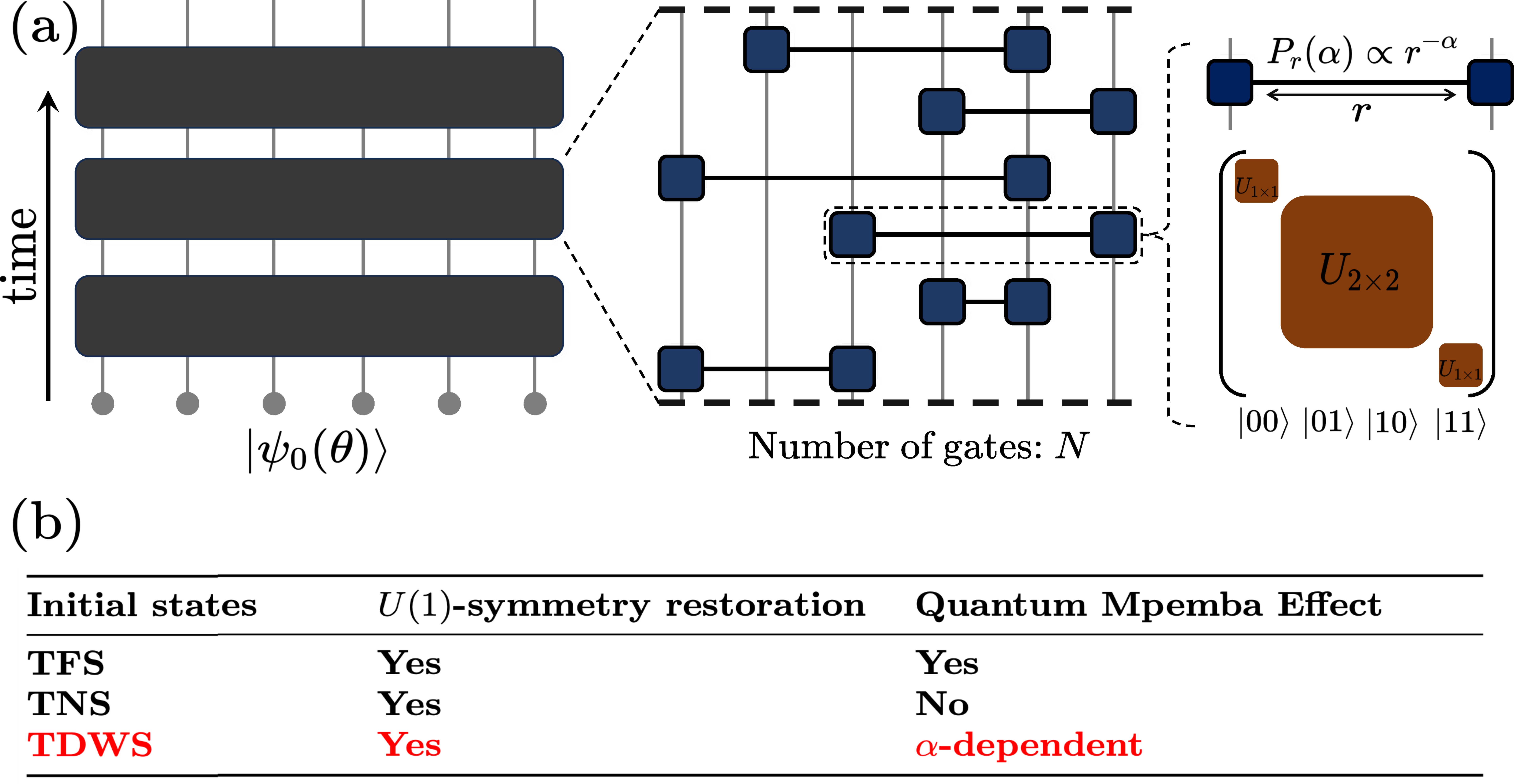} 
\caption{(a) Schematic of the long-range U(1)-symmetric random quantum circuit.  
Starting from the tilted product state $\ket{\psi_0(\theta)}$, we apply layered U(1)-preserving two-qubit gates with power-law range $P_{r_{ij}}(\alpha)\!\propto\!r^{-\alpha}$.
The number of active two-qubit gates per layer equals to the system size $N$. The evolution of the circuit is given by Eq.~\eqref{eq:evolution}. (b) Summary of key dynamical features for three representative initial states, TFS, TNS, and TDWS: whether U(1) symmetry is restored at long-time limit, whether the QME occurs, and whether the presence of QME is $\alpha$-dependent.}
\label{fig:circuit}
\end{figure}

In this work, we employ long-ranged U(1)-symmetric random unitary circuits (RUCs)~\cite{Richter_2023,longRUC1,longRUC2} as a minimal model for long-range quantum thermalized systems, to investigate the QME.
In these circuits, the two-qubit gates are sampled from a distance-dependent distribution [see Fig.~\ref{fig:circuit}(a)], which directly entangles distant degrees of freedom, thereby violating the Lieb-Robinson limitation~\cite{lucas2,lucas3,Fisher_2023,Richter_2023} and enabling faster thermalization. By tuning the interaction exponent $\alpha$,
one can interpolate continuously between diffusive, ballistic, and superdiffusive transport~\cite{lucas0,lucas1,Richter_2023}.
Such circuits faithfully reproduce the key scrambling behavior of power-law interacting Hamiltonian systems~\cite{Richter_2023,PhysRevLett.128.010604}. 

Specifically, we investigate the QME in long-ranged, U(1)-symmetric RUCs on a 1D open chain. To probe symmetry restoration, we employ the annealed R\'enyi-2 entanglement asymmetry for a contiguous subsystem, computed via a two-copy tensor-network construction. We start from three tilted product states: tilted ferromagnetic state (TFS), tilted
N\'eel state (TNS), and tilted ferromagnetic state with a central domain wall (TDWS). We find that the QME is always present for TFS (all $\alpha$); it is absent for TNS; and it appears for TDWS only when the circuit is short-ranged (effectively large $\alpha$). 

In the QME regime, the Mpemba time $t_{\rm M}$, extracted from the crossing of the R\'enyi-2 entanglement asymmetry, scales with the subsystem size as $t_{\rm M}\sim N_A^{z}$ with the dynamical exponent $z=\min(\alpha-1,2)$, extending the short-range scaling $t_{\rm M}\sim N_A^2$~\cite{liu1,xhek}.  We identify a general criterion for the presence of the QME in long-range RUCs: it is governed not only by the U(1) charge-sector structure of the initial states, but also by the long-range-interaction-dependent rates of information and charge propagation in space. These results are ready to be verified on digital quantum simulators, which have recently been shown to be increasingly reliable in dynamically evolving a wide range of Hamiltonians~\cite{Koh_2022,Koh_2024,Tianqi,shenruizhe,digital1,digital2,digital3}.

This paper is organized as follows. In Sec.~\ref{model} we introduce the long-range U(1)-symmetric RUCs and describe the numerical simulation methods used throughout. In Sec.~\ref{results} we present the symmetry-restoration dynamics for different initial states, identify when the QME is present or absent, and, whenever QME occurs, analyze the scaling of the Mpemba time with the subsystem size for various tilt angles. 
In Sec.~\ref{discussion}, we discuss a general criterion of QME in long-ranged circuit. Finally, Sec.~\ref{conclusion} summarizes the main results and outlines possible directions for future work. Additional numerical data are collected in the Appendix.

\section{Model and Probe}\label{model}
\subsection{Long-ranged U(1) symmetric RUCs}
\begin{figure*}[bt]
\hspace*{-0.98\textwidth}
    \centering
    \includegraphics[width=0.98\linewidth]{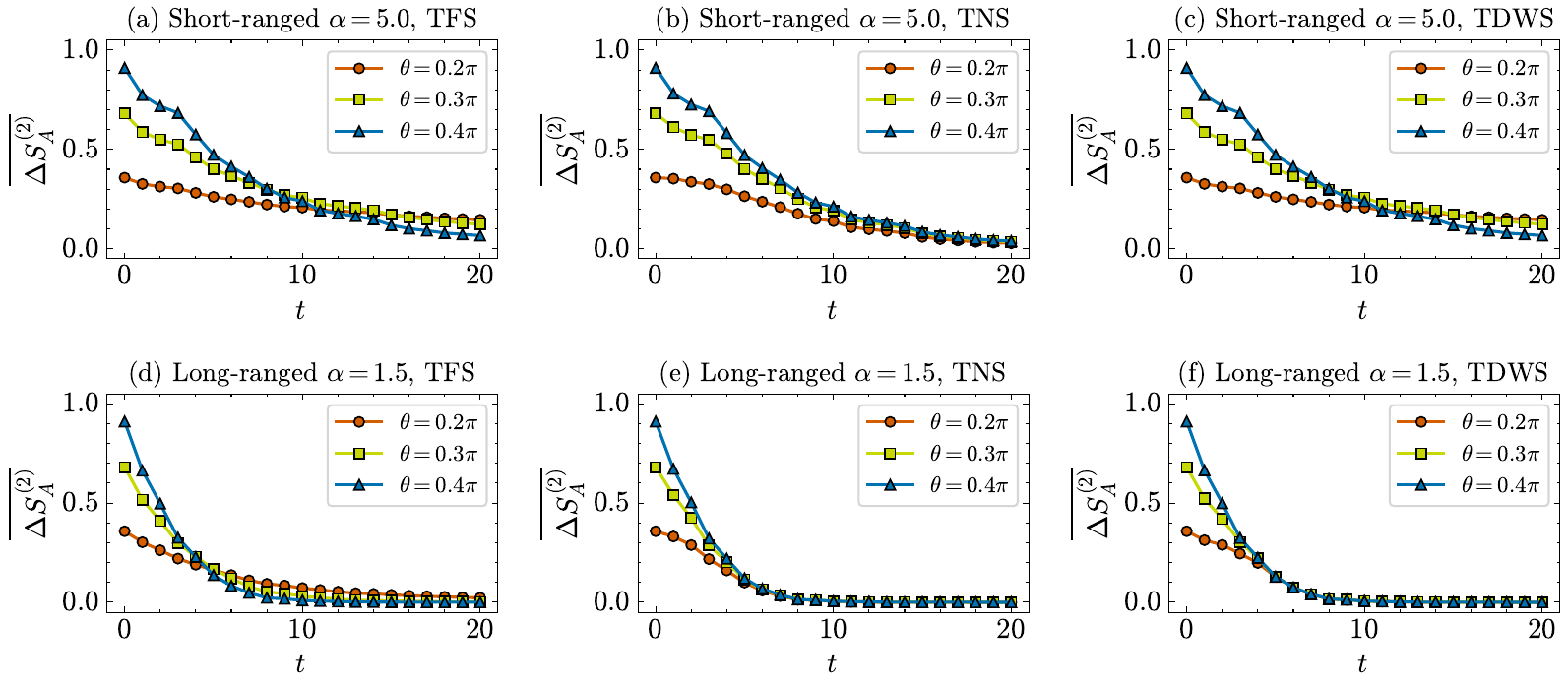}
    \caption{Entanglement asymmetry dynamics for TFS/TNS/TDWS: short- vs. long-range interactions. $\overline{\Delta S_{A}^{(2)}}(t)$ for a subsystem of size $N_{A}=2$ chosen at the left edge of a chain with total length $N=48$. 
    The main panels show numerical results obtained using the replica tensor-network method, averaged over $50$ sampling realizations. 
    Panels (a,d), (b,e), and (c,f) correspond to TFS, TNS, and TDWS initial states, respectively. 
    The upper row (a–c) is for a more short-ranged interaction with exponent $\alpha=5.0$, while the lower row (d–f) is for a more long-ranged interaction with $\alpha=1.5$. 
    (See Appendix~\ref{AppendixB} for additional numerical results.)
    }\label{fig:EA}
\end{figure*}
We consider a one-dimensional open chain with $N$ qubits. The local Hilbert space on site $i$ is $\mathcal{H}_i \cong \mathbb{C}^2$, and the global Hilbert space is $\mathcal{H}\!=\!\bigotimes_{i=1}^N \mathcal{H}_i$. On each site we define a local U(1) charge operator $Q_i\!\equiv\!Z_i/2$ where $Z_i=\sigma_i^z$ is the Pauli matrix at sit $i$, with eigenvalues $q_i\!=\!\pm \tfrac12$. The conserved global charge is $Q\!\equiv\!\sum_{i=1}^N Q_i\!=\! \sum_{i=1}^N Z_i/2$, with eigenvalues denoted by $q$. The global Hilbert space decomposes into charge sectors as $\mathcal{H}\!=\!\bigoplus_q \mathcal{H}_q$, and we denote the dimension of each charge sector by $d_q\!\equiv\!\dim \mathcal{H}_q$. Given a subsystem $A \!\subset\!\{1,\dots,N\}$ of size $N_A\!\equiv\!|A|$, we define the subsystem charge $Q_A\!\equiv\!\sum_{i \in A} Q_i$, with eigenvalues $q_A$ and corresponding subspaces $\mathcal{H}_{A,q_A}$ of dimension $d_{q_A}\!\equiv\!\dim \mathcal{H}_{A,q_A}$.

Time evolution proceeds in discrete layers $t$. In each layer we apply $N$ two-qubit gates $U_{ij}$ acting on pairs of sites $(i,j)$. The distance between two sites is $r_{ij}\!\equiv\!|i-j|$. In each layer, every pair $(i,j)$ is drawn independently from a distance-dependent probability distribution
\begin{align}
P_{r_{ij}}(\alpha) \propto r_{ij}^{-\alpha}, \qquad r_{ij} = 1, 2,\cdots, N-1,
\end{align}
so that $\alpha\!\to\!\infty$ reproduces a short-range circuit dominated by nearest-neighbor gates, while $\alpha\!\to\!0$ approaches an all-to-all circuit.

Each two-qubit gate preserves the local charge $Q_i + Q_j$, i.e.\ $[U_{ij},\, Q_i+Q_j]\!=\!0$.
We denote $d^{(2)}_{q_{\mathrm{loc}}}$ to distinguish these two-site sector dimensions from the global sector dimensions $d_q$. In this decomposition the two-qubit gate can be written as
\begin{align}
U_{ij}\!=\!\bigoplus_{q_{\mathrm{loc}}} U_{ij}^{(q_{\mathrm{loc}})}=
\begin{pmatrix}
U_{1\times 1} & 0 & 0 \\
0 & U_{2\times 2} & 0 \\
0 & 0 & U_{1\times 1}
\end{pmatrix},
\end{align}
where $U_{ij}^{(q_{\mathrm{loc}})}\!\in \!U\big(d^{(2)}_{q_{\mathrm{loc}}}\big)$ and each block $U_{ij}^{(q_{\mathrm{loc}})}$ is independently drawn from the Haar measure on $U(d^{(2)}_{q_{\mathrm{loc}}})$. After $t$ layers, the evolved state is
\begin{align}
    \ket{\psi(t)} = U_t\ket{\psi_0(\theta)}, \qquad 
    U_t = \prod_{\tau=1}^t \prod_{(i,j)\in E_\tau} U_{ij},\label{eq:evolution}
\end{align}
where $E_\tau\!\subset\!\{ (i,j)\vert 1\leq i < j \leq N \}$ denotes the set of long-range pairs chosen at layer $\tau$. 

The random choice of pairs in each layer and in each gate is independent; averages over pair configurations are denoted by an overline $\overline{(\bullet)}$. On the other hand, we denote Haar averages by $\mathbb{E}_{\mathrm{Haar}}[\bullet]$. Notably, the model has two separate sources of randomness: (i) random gate positions and (ii) Haar-random gates.

The initial states are  tilted product states parametrized by an angle $\theta\!\in\!(0,\pi/2]$. The site-wise rotation is defined as $  R_y(\theta)\!=\!\exp\left(- i \frac{\theta}{2} \sum_{i=1}^N Y_i\right)$. We consider three types of initially tilted product states: TFS, TNS, and TDWS, defined as
\begin{eqnarray}
    \ket{\psi_{0}(\theta)} = 
    \begin{cases}  
        R_y(\theta)\, \ket{0}^{\otimes N}, 
        & \text{TFS} \\[6pt]  
        R_y(\theta)\, \big(\ket{01}\big)^{\otimes N/2}, 
        & \text{TNS} \\[6pt]  
        R_y(\theta)\, \big(\ket{0}^{\otimes N/2} \otimes \ket{1}^{\otimes N/2}\big), 
        & \text{TDWS}  
    \end{cases}
    \label{eq:init_states}
\end{eqnarray}

\subsection{Entanglement asymmetry and QME}

In this work, we use the entanglement asymmetry~\cite{Ares2023first, EA7, EA6, EA5, EA4, EA3, EA2, EA1, EA8} as a probe to characterize the strength of U(1) symmetry breaking. 
For the subsystem $A$ with time-evolved reduced density matrix
\begin{align}
\rho_A(t)\equiv{\rm Tr}_{\bar{A}}(U_t\ket{\psi_0(\theta)}\bra{\psi_0(\theta)}U_t^{\dagger}),
\end{align}
where $\bar{A}$ is the complement of $A$, its U(1)-symmetrized version is defined as
\begin{align}
\rho_{A,Q_A}(t)\equiv\sum_{q_A} \Pi_{q_A} \rho_A(t)\Pi_{q_A},
\end{align}
where $\Pi_{q_A}$ is a projector onto the $Q_A$-charge sector $q_A$. The annealed purities are $P_A^{(2)}(t)\!\equiv\!\mathbb{E}_{\rm Haar}[{\rm Tr}\big(\rho_A(t)^2)]$ and $P_{A,Q_A}^{(2)}(t)\!\equiv\!\mathbb{E}_{\rm Haar}[{\rm Tr}\big(\rho_{A,Q_A}(t)^2\big)]$  by averaging random realizations of the gates at fixed spacetime positions, and the annealed~\cite{annealedEA} R\'enyi-2 entanglement asymmetry is $\Delta S_A^{(2)}(t)\!\equiv\!-\log[{P_{A,Q_A}^{(2)}(t)}/{P_A^{(2)}(t)}]\!\ge\!0$, with equality iff $\rho_A(t)\!=\! \rho_{A,Q_A}(t)$, i.e., the subsystem $A$ is fully U(1)-symmetric. 

Then, by averaging over random gate positions, we define 
$\overline{P_A^{(2)}}(t)$ and $\overline{P_{A,Q_A}^{(2)}}(t)$ and the corresponding annealed entanglement asymmetry
\begin{align}
\overline{\Delta S_A^{(2)}}(t) &= -\log\frac{\overline{P_{A,Q_A}^{(2)}}(t)}{\overline{P_A^{(2)}}(t)}.\label{eq:EA0}    
\end{align}
Given two quenches with tilts $\theta_1\!<\!\theta_2$ and initial ordering $\overline{\Delta S_{A}^{(2)}}(\theta_1,0)\!<\!\overline{\Delta S_{A}^{(2)}}(\theta_2,0)$, we say that the QME occurs if there exists a Mpemba time $t_{\rm M}\!>\!0$ such that $\overline{\Delta S_A^{(2)}}(\theta_1, t)>\overline{\Delta S_A^{(2)}}(\theta_2, t)$ for all $t\!>\!t_{\rm M}$, i.e., the entanglement asymmetry order-reversal time. This means that a more strongly U(1)-symmetry-broken initial state relaxes faster. Below, we numerically calculate the symmetry-restoration dynamics for TFS, TNS, and TDWS across $\alpha$ and initial tilt angles $\theta$, and quantify the scaling of the Mpemba time $t_{\rm M}$. 

\section{Results\label{results}}
In this work, we obtain our results numerically using replica tensor-network methods and exact diagonalization. All numerical calculations were carried out with the \texttt{ITensor} library~\cite{itensor1,itensor2} and \texttt{TensorCircuit-NG}~\cite{Tensorcircuit}. See Appendix~\ref{AppendixA} for details of the numerical methods, and Appendices~\ref{AppendixB} and~\ref{AppendixC} for additional numerical results and benchmarks.

\subsection{Entanglement-asymmetry dynamics}
We first numerically calculate the circuit-averaged annealed R\'enyi-2 entanglement asymmetry
$\overline{\Delta S_A^{(2)}}(t)$ for an edge subsystem of size $N_A=2$ in a chain of length $N=48$ with open boundaries, averaged over $50$ circuit realizations [Fig.~\ref{fig:EA}]. The upper row, Figs.~\ref{fig:EA}(a)-(c), corresponds to short-range interactions with exponent $\alpha\!=\!5.0$, while the lower row, Figs.~\ref{fig:EA}(d)-(f), shows more long-ranged interactions with $\alpha\!=\!1.5$. Each column displays a different class of tilted product states, TFS, TNS, and TDWS, and in each panel we compare three representative tilts $\theta\in\{0.2\pi,0.3\pi,0.4\pi\}$.

For TFS [Figs.~\ref{fig:EA}(a),(d)], we find a clear QME: although larger tilts correspond to more asymmetric initial states, the curves for $\theta\!=\!0.4\pi$, $\theta\!=\!0.3\pi$ and $\theta\!=\!0.2\pi$ cross, respectively, so that the initially more asymmetric state restores the subsystem symmetry faster.
The crossings occur earlier for the more long-ranged case $\alpha=1.5$ than that for $\alpha=5.0$, indicating that long-range interactions accelerate the Mpemba-type relaxation. However, TNS initial states [Figs.~\ref{fig:EA}(b),(e)] show no crossings: the ordering of the curves with respect to $\theta$ is preserved throughout the evolution, and no QME is observed. More interestingly, TDWS states [Figs.~\ref{fig:EA}(c),(f)] display an intermediate, interaction-dependent behavior. For short-range interactions, $\alpha\!=\!5.0$, the curves crossed each other, signaling a QME similar to TFS; however, for $\alpha\!=\!1.5$, the curves stayed ordered and the QME was absent. 

In Appendix~\ref{AppendixB}, we have provided additional data for $\alpha\!\in\!\{4,3,2.5,2.0,0.5,0\}$, to further confirm the above conclusion for QME, i.e., (i) robust across interaction ranges for TFS, (ii) absent for TNS, and (iii) present for TDWS only in a finite window of sufficiently short-ranged interactions. These results suggest that for TDWS, the exponent $\alpha$ separating the regimes with and without QME is roughly $\alpha_c\approx 2$.

\subsection{Scalings of Mpemba time}
\begin{figure}[bt]
\hspace*{-0.5\textwidth}
\centering
\includegraphics[width=0.99\linewidth]{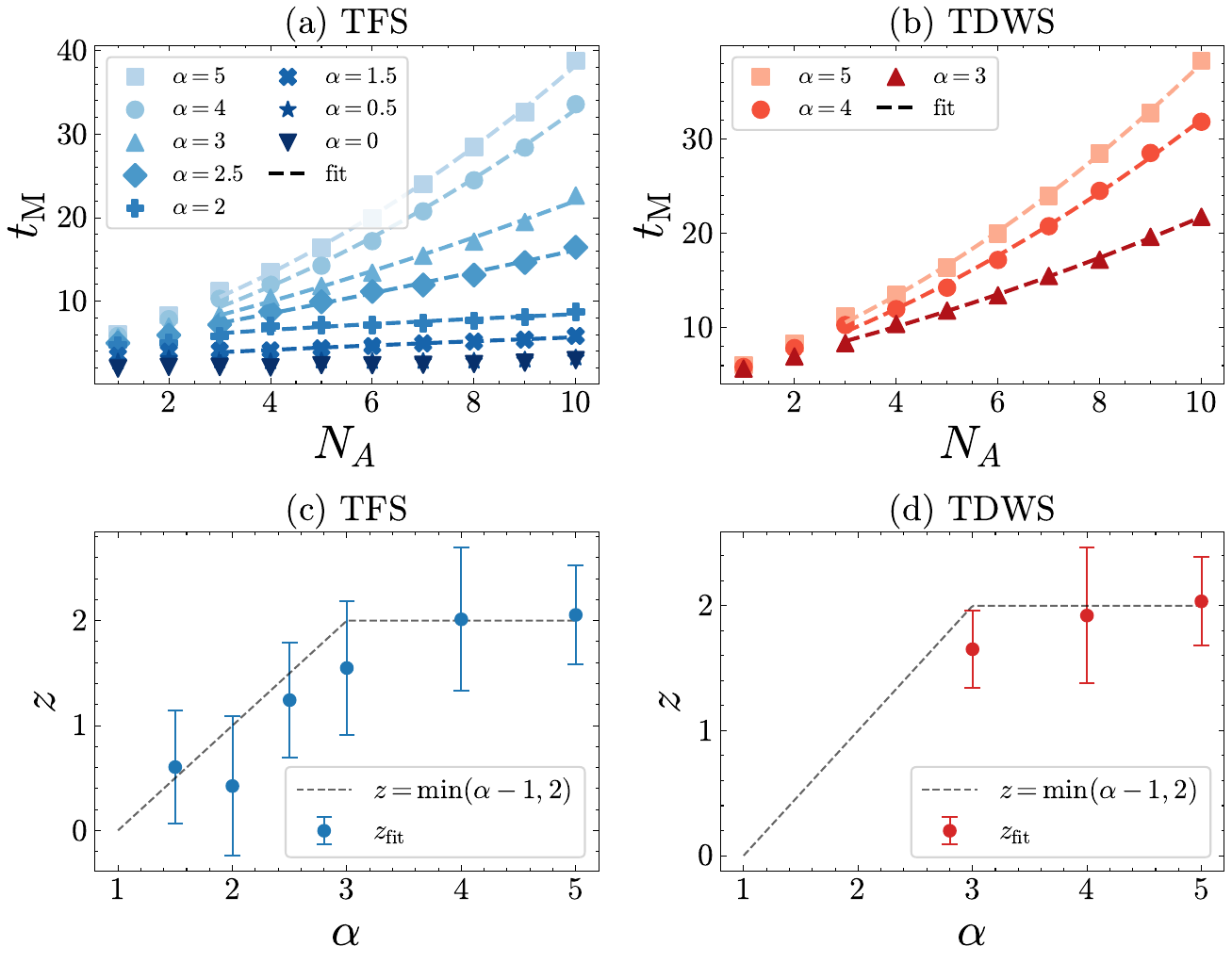} 
\caption{(a)-(b) Mpemba time $t_{\rm M}$ versus subsystem size $N_A$ for TFS and TDWS initial states with $\theta_1=0.3\pi$, $\theta_2=0.4\pi$, chain lengths $N=48$, and $N_A\!=\!1,2,\dots,10$. Symbols denote replica tensor-network simulation results; dashed lines are fits to $t_{\rm M}\!=\!b\,(N_A - x_0)^{z} + c$, performed only for $N_A\geq3$ and $\alpha\geq1$. (c)-(d) Fitted scaling dynamical exponent $z_{\mathrm{fit}}$ as a function of the long-ranged interaction exponent $\alpha$, compared with the theoretical prediction $z\!=\!\min(\alpha-1,2)$~\cite{longRUC2,Richter_2023,PhysRevX.8.031057}. Fits are performed only for $\alpha\geq1$ in TFS, and for $\alpha \geq3$ in TDWS.}\label{fig:theory}
\end{figure}
We have demonstrated the QME with fixed subsystem size $N_{A}=2$. Next we investigate the dependence of Mpemba time on the subsystem size. 
We define the Mpemba time $t_{\rm M}$ as the first crossing between the entanglement-asymmetry curves for $\theta_{1}$ and $\theta_{2}$. Here we choose $\theta_1=0.3\pi$ and $\theta_2=0.4\pi$. Fig.~\ref{fig:theory} shows the circuit-averaged $t_{\rm M}$ as a function of subsystem size $N_A$ for chains with fixed total system size $N=48$. We focus on the cases where the QME is present: TFS for a broad range of $\alpha$ [Figs.~\ref{fig:theory}(a) and (c)] and TDWS for short-ranged ($\alpha\geq 3$) interactions [Figs.~\ref{fig:theory}(b) and (d)].

For TFS [Fig.~\ref{fig:theory}(a)], $t_{\rm M}$ increases monotonically with $N_A$ for all $\alpha$. At large exponents (short-ranged), the growth is strongly curved and consistent with a quadratic scaling, $t_{\rm M}\sim N_A^2$, as expected for diffusive charge transport in short-range circuits. As $\alpha$ is reduced and the interactions become more long-ranged, the $t_{\rm M}(N_A)$ curves straighten and approach an almost linear dependence, consistent with earlier crossings in the long-range regime.
Power-law fits of the form $t_{\rm M}=b(N_A-x_0)^z+c$ yield dynamical exponents $z_{\rm fit}$ that track the long-range entanglement-spreading prediction $z$ [Fig.~\ref{fig:theory}(c)], thus tying the Mpemba timescale directly to the underlying transport of entanglement~\cite{longRUC2,Richter_2023,PhysRevX.8.031057}.

For TDWS, a similar scaling is observed but only for those $\alpha$ where the QME exists. For $\alpha=4$ and $5$, $t_{\rm M}$ again grows with $N_A$ and is well described by power-law fits with exponents $z_{\rm fit}$ compatible with $z=\min(\alpha-1,2)$ [Figs.~\ref{fig:theory}(b),(d)]. For smaller $\alpha$, no crossings occur and $t_{\rm M}$ is not defined.
In Fig.~\ref{fig:theory}(d), the short-ranged behavior suggests that within the regime where QME is visible, it satisfies the dynamical exponent formula $z$.

These results show that QME is highly sensitive to both the structure of the initial state and the interaction range, and its characteristic timescale inherits the crossover from ballistic to superdiffusive entanglement spreading in long-ranged RUCs.

\section{Discussion}\label{discussion}
\subsection{QME in long-ranged U(1)-symmetric RUCs}
We now discuss how long-ranged circuits affect symmetry-restoration dynamics and the emergence of the QME. For any initial state $\ket{\psi_0(\theta)}$ we denote by $p_{q_A}(\theta)=\mathrm{Tr}[\Pi_{q_A}\rho_A(0)]$ the weight of its reduced density matrix $\rho_A(0)$ in the subsystem charge sector $q_A$.
In the long-time limit, and for any fixed global charge $Q$, the dynamics scrambles states within each charge sector to a U(1)-symmetric Haar-random state. As a result, long-time averaged observables on $A$---such as the R\'enyi-2 entanglement asymmetry---depend only on the global charge-sector weights and are essentially insensitive to microscopic circuit details: brickwork and power-law circuits reach the same steady state, consistent with Ref.~\cite{liu1}. For initial states whose U(1) symmetry is fully restored in the thermodynamic limit---TNS and TDWS with $|A|<N/2$ at any $\theta$, and TFS within a fixed charge sector, the late-time value is well described by a block-diagonal U(1)-symmetric Haar ensemble and is independent of $\alpha$. By contrast, for slightly tilted TFS at finite $N$, subsystem symmetry is not fully restored and a finite symmetry-breaking plateau survives, in agreement with Refs.~\cite{liu1,xhek}.

The QME, instead, probes the approach to this equilibrium. For a charge-conserving circuit at fixed exponent $\alpha$, relaxation within a given global charge sector explores the subsystem sectors $\mathcal{H}_{A,q_A}$ of dimensions $d_{q_A}$. The relaxation rate is controlled by how the initial state distributes its weights $p_{q_A}(\theta)$ over sectors of different sizes, together with how efficiently the conserved charge can propagate through the system for the chosen $\alpha$. If most of the weight lies in extreme sectors with $|q_A|\approx N_A/2$, the number of accessible microstates is very small and thermalization is strongly constrained, irrespective of the propagation speed. If the weight is instead concentrated in the central sectors with small $|q_A|$ and large $d_{q_A}$, random gates can efficiently mix states within those sectors and, provided charge and entanglement spread sufficiently fast, the approach to equilibrium is greatly accelerated. When propagation is mainly mediated by short-range gates, $t_{\rm M}$ grows rapidly with subsystem size, whereas in long-ranged circuits with small $\alpha$ strong long-distance gates redistribute charge and entanglement across $A$ within only a few layers, substantially reducing $t_{\rm M}$. QME in long-ranged chaotic U(1)-conserving systems thus originates from the (generally non-monotonic) dependence of the overlap between $\{p_{q_A}(\theta)\}$ and $\{d_{q_A}\}$ on the tilt angle $\theta$, combined with the propagation rate set by $\alpha$.

This perspective makes the contrast between TFS and TNS transparent. For TFS, $\theta$ mainly controls how the initial state populates the $q_A$ sectors: small $\theta$ puts most weight in extreme sectors with large $|q_A|$ and small $d_{q_A}$, leading to slow relaxation, whereas larger $\theta$ shifts weight towards central sectors with small $|q_A|$ and large $d_{q_A}$, leading to fast relaxation. Because the value of $\theta$ that minimizes the initial asymmetry does not coincide with the one that best populates the large-$d_{q_A}$ sectors, TFS exhibits a robust QME. For TNS, by contrast, the weight over $q_A$ sectors remains concentrated near the central, large-dimension sectors over a wide range of $\theta$; changing $\theta$ mostly rearranges configurations within these sectors. The relaxation time then varies almost monotonically with $\theta$, and TNS does not exhibit a visible QME in the regimes we study.

TDWS lies between these two limits. It can be viewed as two large ferromagnetic domains connected by a sharp interface at the chain center, followed by an overall tilt by $\theta$. For large $\alpha$, where the dynamics is effectively short-ranged, charge must cross the domain wall through local gates, so the macroscopic charge imbalance between the two domains relaxes slowly. Under the fixed global charge constraint, such a configuration is effectively obtained by gluing together two TFS-like regions, and the sector weights $\{p_{q_A}(\theta)\}$ therefore change with $\theta$ in close analogy to TFS. In this regime TDWS exhibits a clear QME.

When $\alpha$ is decreased and the circuit enters the strongly long-ranged regime, the power-law distribution $P_{r_{ij}}(\alpha)\propto r_{ij}^{-\alpha}$ allows many distant sites to be coupled within a single layer, and the domain-wall structure is rapidly washed out. For each fixed global charge $Q$, the state quickly approaches, within the corresponding sector $\mathcal{H}_Q$, a spatially random configuration that is well approximated by a U(1)-symmetric Haar-random state. In this regime locality is effectively absent, and TDWS behaves similarly to TNS: the weight over $q_A$ sectors is dominated by the central, large-dimension sectors for all $\theta$, and changing $\theta$ mainly rearranges configurations within these sectors. Consequently, TDWS no longer exhibits a visible QME in long-ranged circuits with small $\alpha$.

In summary, the long-time behavior of the entanglement asymmetry is fixed by global properties of the system as the U(1) symmetry and the charge-sector structure of the initial state, so states like TNS and TDWS, which share the same global charge sector weights, have essentially identical late-time behavior but differ from TFS. By contrast, the early-time dynamics and the presence of a QME are controlled by how local the dynamics is: when gates are predominantly short-ranged, spatial structure and sector populations evolve slowly and TFS- and TDWS-like states can exhibit a pronounced QME, whereas in the strongly long-ranged regime locality is effectively absent and TDWS behaves similarly to TNS without a visible QME. Thus, the interplay between the global sector structure and the degree of locality in the dynamics determines whether different initial states can exhibit a QME in long-ranged U(1)-symmetric RUCs, and how pronounced this effect is.

\subsection{Comparison with other closed quantum long-ranged systems}
Previous works on the QME in closed quantum long-ranged Hamiltonian systems have mainly focused on the dynamical restoration of U(1) symmetry. Ref.~\cite{14} instead attributes the QME to the rate at which zero-momentum spin-wave fluctuations, generated after a quench, melt an initially ferromagnetic order; in this picture, the role of long-range interactions is encoded in the generation and propagation of these spin waves. On the other hand, Ref.~\cite{Matthew}, exemplified by long-range XYZ spin chains, tuning the range of interactions can induce or suppress spontaneous symmetry breaking, so that the QME is observed only in the disordered regime while it is terminated in the symmetry-broken phase. Taken together, Refs.~\cite{14} and \cite{Matthew} identify the key conditions for a QME in long-range Hamiltonian spin systems: long-range couplings must enhance the restoration of a broken continuous symmetry after a quench by accelerating the growth and propagation of collective fluctuations.

By contrast, our work on U(1)-symmetric long-range RUCs reveals a complementary yet consistent organizing principle for the QME in long-ranged systems. Our core finding, in long-ranged U(1) symmetric RUCs, is that the presence or absence of a QME is jointly dictated by these two ingredients: the global charge-sector structure of the initial states and the speed of information and charge propagation. This perspective is complementary to, and consistent with, Refs.~\cite{14} and \cite{Matthew}.

\section{Conclusion and outlook\label{conclusion}}
In this work, we investigate the QME in long-ranged, U(1)-symmetric random unitary circuits. The circuits act on a one-dimensional open chain and realize a tunable interaction range via a distance-dependent two-qubit sampling probability $P_{r_{ij}}(\alpha)\!\propto\!r_{ij}^{-\alpha}$. As a probe of symmetry breaking, we used the annealed R\'enyi-2 entanglement asymmetry, which we evaluated for a contiguous subsystem by the replica tensor-network method formulated in the two-copy Liouville space. we investigate
three tilted product states introduced earlier: the TFS, the TNS, and the TDWS. Our simulations exhibit the following pattern: the QME is robust for the TFS for all $\alpha$; by contrast, it is absent for the TNS; and, moreover, it occurs for the TDWS only when the circuit is effectively short-ranged (effectively large $\alpha$). In the QME cases, the circuit-averaged Mpemba time $t_{\rm M}$ extracted from the crossings of the annealed R\'enyi-2 entanglement asymmetry fits a power-law form, and the fitted exponent agrees with the long-ranged diffusion exponent $z\!=\!\min(\alpha-1,2)$. Thus the 
short-range law $t_{\rm M}\!\sim\!N_A^{2}$~\cite{liu1,xhek} is continuously extended to the fractional-diffusion regime realized by power-law gates. 

Additionally, the global U(1) symmetry and its charge-sector structure of the initial states fix the universal long-time plateau, which is captured by a global U(1)-symmetric Haar ensemble and is independent of the interaction range. The power-law exponent $\alpha$ sets the strength of long-range gates and thus the rate of information and charge spreading, which controls how fast spatial structure in the initial state is erased and the sector weights are effectively redistributed. We find that the interplay between the charge-sector structure of the initial states and the propagation speed determines whether a given initial state exhibits a QME in short- and long-range RUCs, and how prominently the effect appears. Compared with previous Hamiltonian studies~\cite{14,Matthew} that emphasize symmetry restoration via collective fluctuations and the stabilization of symmetry-broken phases, our work provides a complementary organizing principle in which the QME is governed by U(1) charge-sector structure of the initial states together with long-range-interaction-controlled transport.

Our work deepens the understanding of universal mechanisms in quantum nonequilibrium dynamics and QME, and thereby opens avenues for controlling complex quantum systems. Future studies may investigate whether charged dual-unitary circuits~\cite{Bertini} with solvable initial states can also exhibit fast thermalization in the absence of a QME. It would also be highly interesting to explore the fate of the QME in models where the thermalization speed is tuned or controlled by measurements~\cite{skinner,lyd,lsM0,lsM1,lsM2,lsM3,lsM4}.

\hspace{3cm}
\begin{acknowledgements}
\noindent
We thank He-Ran Wang and Sheng Yang for helpful discussions. J.-X.\ Zhong was supported by the National Natural Science Foundation of China (Grant Nos.\ 12374046 and 11874316), the Shanghai Science and Technology Innovation Action Plan (Grant No.\ 24LZ1400800), the National Basic Research Program of China (Grant No.\ 2015CB921103), and the Program for Changjiang Scholars and Innovative Research Teams in Universities (Grant No.\ IRT13093). S.\ L.'s work at Princeton University was supported by the Gordon and Betty Moore Foundation (Grant No.\ GBMF8685 toward the Princeton theory program) and its EPiQS Initiative (Grant No.\ GBMF11070); the Office of Naval Research (ONR; Grant No.\ N00014-20-1-2303); the Global Collaborative Network Grant at Princeton University; the Simons Investigator Grant (No.\ 404513); the U.S.--Israel Binational Science Foundation (BSF; No.\ 2018226); the NSF MRSEC program (DMR-2011750); the Simons Collaboration on New Frontiers in Superconductivity; and the Schmidt Foundation at Princeton University. S.-X.\ Zhang acknowledges support from the Quantum Science and Technology-National Science and Technology Major Project (Grant No.\ 2024ZD0301700) and a start-up grant at IOP-CAS. C. H. Lee acknowledges support from the Ministry of Education, Singapore (MOE Tier-II Award No.\ T2EP50224-0021). H.-Z.\ Li is supported by a China Scholarship Council (CSC) Scholarship.
\end{acknowledgements}

%-------------refs---------------------------
\bigskip
\bibliography{refs}

%%%%%%%%%%%%%%%%% APPENDIX %%%%%%%%%%%%%%%%%%%

\clearpage
\newpage
\onecolumngrid

\appendix
\section{Replica tensor-network simulation}
\label{AppendixA}
In this section, we elaborate the replica tensor-network method for long-ranged, U(1)-symmetric random quantum circuits~\cite{xhek}. This enables efficient evaluation of the annealed entanglement asymmetry for large systems. Please refer to the main text for the notation of the long-ranged U(1)-symmetric RUCs.

To analyze the R\'enyi-2 purity we work in Liouville space. For an operator $X$ acting on $\mathcal{H}$ with matrix elements $X_{xy}\!=\!\bra{x}X\ket{y}$ in some orthonormal basis $\{\ket{x}\}$, we define the vectorization $| X \rrangle\!\equiv \!\sum_{x,y} X_{xy} \, \ket{x} \otimes \ket{y}$.
The Hilbert--Schmidt inner product $\mathrm{Tr}(Y^\dagger X)$ is then represented as
$\mathrm{Tr}(Y^\dagger X)\!= \!\llangle Y \vert X \rrangle$.
A unitary channel $X \mapsto UXU^\dagger$ is represented as $\vert X \rrangle \mapsto (U \otimes U^*) \vert X \rrangle$.
For the R\'enyi-2 quantities, we work with two replicas, so the Liouville space is that of operators on $\mathcal{H}^{\otimes 2}$, and each physical site carries four legs: replica $1/2$ and ket/bra. For a given circuit realization, the two-replica superoperator after $t$ layers is
\begin{align}
  \mathcal{U}_t &\equiv \big(U_t \otimes U_t^*\big)^{\otimes 2}.
\end{align}

The global pure-state density matrix is $\rho(t)\!=\!\ket{\psi(t)}\bra{\psi(t)}$, and the reduced density matrix for subsystem $A$ is $\rho_A(t)\!= \!\mathrm{Tr}_{\bar{A}}\rho(t)$. The R\'enyi-2 purity is $P_A^{(2)}(t)\!\equiv\!\mathrm{Tr}\left[\rho_A(t)^2\right]$.
Using the standard swap trick, it can be written as $P_A^{(2)}(t)\!=\!\mathrm{Tr}\left[F_A \, \rho(t) \otimes \rho(t)\right]$, where $F_A$ swaps the two replicas on subsystem $A$ and acts as the identity on $\bar{A}$. In the two-replica Liouville representation this becomes
\begin{align}
  P_A^{(2)}(t)
  &= \llangle F_A \vert \, \mathcal{U}_t\,\vert \rho_0^{\otimes 2} \rrangle,
\end{align}
where $\vert \rho_0^{\otimes 2} \rrangle$ is the vectorization of $\rho_0 \otimes \rho_0$, and $\llangle F_A \vert$ is the corresponding boundary vector. The charge-dephased reduced density matrix is $\rho_{A,Q_A}(t) \equiv \mathcal{C}_A(\rho_A(t))$, and its R\'enyi-2 purity is $P_{A,Q_A}^{(2)}(t)\!\equiv\!\mathrm{Tr}[\rho_{A,Q_A}(t)^2]$. 

At each site $i$, the four-leg replica space is $\mathcal{K}_i\!\equiv\!(\mathbb{C}^2)^{\otimes 4}$. We label the four legs as (replica 1, ket), (replica 1, bra), (replica 2, ket), (replica 2, bra). 
For the computational-basis state $\vert \sigma_1, \sigma_2, \sigma_3, \sigma_4 \rrangle_i$ on site $i$ we denote the local charges on the four legs by $q_s\!\equiv\! q(\sigma_s)\!\in\!\{ \pm\tfrac12 \}$ for $s\!=\!1,\cdots,4$. We now introduce two ``replica charges''
% which characterize the local charge of replica 1 and replica 2, respectively, with 
$r,b \in \{\pm \tfrac12\}$, which serve as labels for the local U(1) charges carried by replica $1$ and replica $2$ on site $i$. 
Following Ref.~\cite{xhek}, we define the \emph{parallel} ($+$) and \emph{crossed} ($-$) single-site replica states
\begin{align}
  \vert +,r,b\rrangle_i 
  &\equiv \sum_{\sigma_1,\sigma_2}
      \vert \sigma_1,\sigma_1,\sigma_2,\sigma_2\rrangle_i\,
      \delta_{q(\sigma_1),r}\,\delta_{q(\sigma_2),b},\\
  \vert -,r,b\rrangle_i 
  &\equiv \sum_{\sigma_1,\sigma_2}
      \vert \sigma_1,\sigma_2,\sigma_2,\sigma_1\rrangle_i\,
      \delta_{q(\sigma_1),r}\,\delta_{q(\sigma_2),b}.
\end{align}
For spin-$\tfrac12$ one finds two exact degeneracies,
\begin{align}
\begin{split}
  \vert +,+\tfrac12,+\tfrac12\rrangle_i 
  &= \vert -,+\tfrac12,+\tfrac12\rrangle_i,\\
  \vert +,-\tfrac12,-\tfrac12\rrangle_i 
  &= \vert -,-\tfrac12,-\tfrac12\rrangle_i,
\end{split}
\end{align}
so there are only six linearly independent local replica states.
Equivalently, they span the six-dimensional subspace
\begin{align}
\mathcal{K}_i^{\mathrm{eff}}
  \equiv \mathrm{span}\big\{\,\vert +,r,b\rrangle_i,\;
     \vert -,r,b\rrangle_i
     \;\big|\; r,b=\pm\tfrac12
  \big\},
\end{align}
subject to the identities above.
We then choose an orthonormal basis 
$\{\ket{\mu}_i\}_{\mu=1}^6$ of $\mathcal{K}_i^{\mathrm{eff}}$, 
obtained for instance by orthonormalizing any six linearly independent
vectors in this set, where
the index $\mu$ simply labels these six local replica basis states. All averaged dynamics relevant for purity and entanglement asymmetry can be restricted to the tensor product $\bigotimes_{i=1}^N \mathcal{K}_i^{\mathrm{eff}}$.

We now consider the Haar-averaged action of a single U(1)-symmetric two-qubit gate in the two-replica Liouville space. On the two-site Hilbert space $\mathfrak{h}_{ij}\!\equiv\!\mathcal{H}_i \otimes \mathcal{H}_j$ we define $Q_{ij} \equiv Q_i + Q_j$, whose eigenvalues $q_{\mathrm{loc}}\!\in\!\{-1,0,+1\}$ correspond to the computational basis states $\ket{11}$, $\{\ket{01},\ket{10}\}$, and $\ket{00}$ respectively, with local sector dimensions $  d^{(2)}_{+1}\!=\!1$, $d^{(2)}_0\!=\!2$ and $d^{(2)}_{-1}\!=\!1$. For fixing a pair $(i,j)$, in $\mathfrak{h}_{ij}$ each local charge sector $q_{\rm loc}$ has dimension $d^{(2)}_{q_{\rm loc}}$. For sectors $q_1$ and $q_2$ we choose orthonormal bases $\{\ket{x_{q_1}}\} \subset \mathfrak{h}_{ij,q_1}$ and $\{\ket{x_{q_2}}\} \subset \mathfrak{h}_{ij,q_2}$, and define the two-site invariant vectors
\begin{align}
  \vert I^+_{q_1,q_2}\rrangle 
  &\equiv \sum_{x_1,x_2} 
      \vert x_1,x_1,x_2,x_2\rrangle \otimes \vert x_1,x_1,x_2,x_2\rrangle,\\
  \vert I^-_{q_1,q_2}\rrangle 
  &\equiv \sum_{x_1,x_2} 
      \vert x_1,x_2,x_2,x_1\rrangle \otimes \vert x_1,x_2,x_2,x_1\rrangle.
\end{align}
For fixed $(q_1,q_2)$ these two vectors span a $2$-dimensional subspace that is invariant under the Haar-averaged action of the random gate. Let $G_{ij} \equiv (U_{ij} \otimes U_{ij}^*)^{\otimes 2}$ be the two-site superoperator in two replicas and Liouville space. Using the Weingarten calculus for unitary groups $U(d^{(2)}_q)$ and the block structure imposed by charge conservation, one finds that the Haar-averaged two-site channel can be written solely in terms of the invariant vectors above. The result is
\begin{align}
\begin{split}
  T_{ij}
  &\equiv \mathbb{E}_{\mathrm{Haar}}[G_{ij}]\\
  &=
  \sum_{q_1 \neq q_2}
  \frac{1}{d^{(2)}_{q_1} d^{(2)}_{q_2}}\,
  \Big(
    \vert I^+_{q_1,q_2}\rrangle \llangle I^+_{q_1,q_2}\vert
    +
    \vert I^-_{q_1,q_2}\rrangle\llangle I^-_{q_1,q_2}\vert
  \Big) 
  \\&\quad+
  \sum_{q : d_q^{(2)}>1}
  \frac{1}{\big(d^{(2)}_q\big)^2 - 1}
  \Big[
    \vert I^+_{q,q}\rrangle\llangle I^+_{q,q}\vert
    + \vert I^-_{q,q}\rrangle \llangle I^-_{q,q}\vert
    - \frac{1}{d^{(2)}_q}
      \Big(
        \vert I^+_{q,q}\rrangle \llangle I^-_{q,q}\vert
        + \vert I^-_{q,q}\rrangle \llangle I^+_{q,q}\vert
      \Big)
  \Big].
\end{split}
\end{align}
Here $T_{ij}$ acts nontrivially only on the replica degrees of freedom on sites $i$ and $j$, and as the identity elsewhere. It is determined entirely by the U(1) charge-sector structure and Weingarten coefficients, and is independent of the distance $r_{ij}$. All Haar randomness has been absorbed into $T_{ij}$.

We now incorporate the second source of randomness, namely that of the random pair selection in each layer. For layer $t$, let $\mathcal{C}_t\!=\!\{(i_m^{(t)}, j_m^{(t)})\}_{m=1}^N$ be the $N$ pairs chosen in that layer. For fixed $\mathcal{C}_t$, the two-replica superoperator is $\mathcal{U}^{(t)}(\mathcal{C}_t) \!\equiv\!\prod_{m=1}^N G_{i_m^{(t)} j_m^{(t)}}$, and its conditional Haar average is $  \mathbb{E}_{\mathrm{Haar}}\big[\mathcal{U}^{(t)}(\mathcal{C}_t) \,\big|\, \mathcal{C}_t\big]\!=\!\prod_{m=1}^N T_{i_m^{(t)} j_m^{(t)}}$.
Averaging this over the random pair configurations using the overline, we obtain $  \overline{\mathcal{U}^{(t)}} \!\equiv\!\overline{\mathbb{E}_{\mathrm{Haar}}\big[\mathcal{U}^{(t)}(\mathcal{C}_t) \,\big|\, \mathcal{C}_t\big]}  \!=\!\overline{\prod_{m=1}^N T_{i_m^{(t)} j_m^{(t)}}}$.
Since the $N$ pairs in a layer are independent draws from the same distribution $P_{r_{ij}}(\alpha)$, this average factorizes: $  \overline{\prod_{m=1}^N T_{i_m^{(t)} j_m^{(t)}}} \!=\!\Big(\overline{T_{ij}}\Big)^N$.
This motivates the definition of the effective per-gate averaged channel $\mathcal{T}_\alpha^{(1)}\!\equiv\!\overline{T_{ij}}\!=\!\sum_{1 \le i < j \le N} P_{r_{ij}}(\alpha)\, T_{ij}$.
Each layer of $N$ gates is thus described, after both Haar and pair averaging, by the superoperator $\overline{\mathcal{U}^{(t)}}\!=\!\big(\mathcal{T}_\alpha^{(1)}\big)^N$. Assuming that different layers are independent and identically distributed, the averaged two-replica evolution after $t$ layers is $\overline{\mathcal{U}_t}\!=\!\big(\mathcal{T}_\alpha^{(1)}\big)^{Nt}$. Substituting this into the expression for the R\'enyi-2 purity, we obtain
\begin{align}
  \overline{P_A^{(2)}(t)}
  &= \llangle F_A \vert \big(\mathcal{T}_\alpha^{(1)}\big)^{Nt} \vert \rho_0^{\otimes 2} \rrangle.
\end{align}
Next, to define the entanglement asymmetry, we introduce the charge-dephasing channel on $A$. The dephasing channel is defined by $\mathcal{C}_A(X)\!\equiv\!\sum_{q_A} \Pi_{q_A} X \Pi_{q_A}$, where $\Pi_{q_A}$ is the projector onto the eigenspace of $Q_A$ with eigenvalue $q_A$. For a subsystem of size $N_A$ consisting of spin-$\tfrac12$ degrees of freedom, there are $N_A+1$ different values of $q_A$. We define $\varphi_k$ for the Fourier representation, defining $\varphi_k\!\equiv\!\frac{2\pi k}{N_A + 1}$ with $k = 0,\dots,N_A$. Using the discrete Fourier representation, we can write $\Pi_{q_A}\!=\!\frac{1}{N_A+1} \sum_{k=0}^{N_A} e^{-i\varphi_k q_A} e^{i\varphi_k Q_A}$. Substituting this into the definition of $\mathcal{C}_A$ and summing over $q_A$ yields
\begin{align}
  \mathcal{C}_A(X) 
  &= \frac{1}{N_A+1} \sum_{k=0}^{N_A} 
    e^{-i\varphi_k Q_A} X e^{+i\varphi_k Q_A}.
\end{align}
The charge-dephased reduced density matrix is $\rho_{A,Q_A}(t) \equiv \mathcal{C}_A(\rho_A(t))$, and its R\'enyi-2 purity is $P_{A,Q_A}^{(2)}(t) \equiv \mathrm{Tr}[\rho_{A,Q_A}(t)^2]$. In the two-replica Liouville representation this can be written as
\begin{align}
  P_{A,Q_A}^{(2)}(t)
  &= \llangle F_A \vert (\mathcal{C}_A \otimes \mathcal{C}_A)\vert \rho(t) \otimes \rho(t)\rrangle.
\end{align}
Since $\mathcal{C}_A$ is a completely positive, trace-preserving map constructed from orthogonal projectors, it is self-adjoint with respect to the Hilbert-Schmidt inner product. It is therefore convenient to move it onto the boundary vector. We define the twisted swap boundary state
\begin{align}
  \vert F_A' \rrangle &\equiv (\mathcal{C}_A \otimes \mathcal{C}_A)\vert F_A \rrangle,
\end{align}
so that
\begin{align}
  P_{A,Q_A}^{(2)}(t)
  &= \llangle F_A' \vert \rho(t) \otimes \rho(t)\rrangle.
\end{align}
Using the twirling form of $\mathcal{C}_A$, one finds that $\vert F_A' \rrangle$ decomposes as an average over a family of $k$-dependent boundary states: $\vert F_A'\rrangle\!=\!\frac{1}{N_A+1} \sum_{k=0}^{N_A} \vert F_A(k) \rrangle$, where each $\vert F_A(k) \rrangle$ is a tensor product of single-site boundary vectors. On sites $i \notin A$, the dephasing acts trivially, so $\vert F^{(i)}(k) \rrangle = \vert F^{(i)} \rrangle$. On sites $i \in A$, the boundary vector can be expressed as a linear combination in the replica-charge basis:
\begin{align}
  \vert F^{(i)}(k) \rrangle 
  &= \sum_{r,b = \pm 1/2} \mathcal{A}_k(r,b) \vert-,r,b\rrangle_i,
\end{align}
with coefficients $\mathcal{A}_k(r,b)\!=\!i^{\,r-b} \tfrac{[\sin(\varphi_k)]^{r+b+1}}{[\cos(\varphi_k)]^{r+b-1}}$. For spin-$\tfrac12$, the states $\vert -,r,b\rrangle_i$ live inside the $6$-dimensional effective space $\mathcal{K}_i^{\mathrm{eff}}$, so each $\vert F^{(i)}(k) \rrangle$ can be expanded in the basis $\{\ket{\mu}_i\}$, and each $\vert F_A(k) \rrangle$ is a matrix product state (MPS)-type boundary state in the effective replica space.

Combining this twisted boundary with the averaged evolution $(\mathcal{T}_\alpha^{(1)})^{Nt}$, we obtain the annealed (Haar and pair averaged) dephased purity
\begin{align}
  \overline{P_{A,Q_A}^{(2)}(t)} 
  &= \frac{1}{N_A+1} \sum_{k=0}^{N_A}
    \llangle F_A(k) \vert \big(\mathcal{T}_\alpha^{(1)}\big)^{Nt}
    \vert \rho_0^{\otimes 2} \rrangle.
\end{align}
Together with
\begin{align}
  \overline{P_A^{(2)}(t)}
  &= \llangle F_A \vert \big(\mathcal{T}_\alpha^{(1)}\big)^{Nt}
    \vert \rho_0^{\otimes 2} \rrangle,
\end{align}
this yields the annealed entanglement asymmetry
\begin{align}
\overline{\Delta S_A^{(2)}}(t) &= -\log\frac{\overline{P_{A,Q_A}^{(2)}}(t)}{\overline{P_A^{(2)}}(t)}.\label{numerics}
\end{align}

Finally, we summarize the numerical implementation of the replicated tensor-network evolution in the above notation. In the replicated Liouville space we denote the global state by $\vert \Psi_t \rrangle$. Formally, $\vert \Psi_t \rrangle$ can be viewed as the result of applying a specific random matrix product operator (MPO), corresponding to one realization of random gates and random pairs, to the initial state $\vert \rho_0^{\otimes 2} \rrangle$; at the level of averaged dynamics, its evolution is described by $(\mathcal{T}_\alpha^{(1)})^{Nt}$. Numerically, we always work in the local basis $\{\ket{\mu}_i\}_{\mu=1}^6$ and adopt the same orthonormal convention as in the definition of these basis states. An overall normalization factor is immaterial for the transfer-tensor evolution and for the purity overlaps, so it is enough to keep the local basis orthonormal at each site. The global state $\vert \Psi_t \rrangle$ is stored as an MPS of local dimension $6$, and a single time step (or layer) of the evolution is implemented as a random MPO, which we denote by $\mathbb{T}_t$. The structure of $\mathbb{T}_t$ corresponds to the tensor network built from the $N$ two-qubit gates in one layer.

To control the growth of the bond dimension in the replicated tensor-network simulation, we monitor the discarded weight after each SVD truncation. For all data shown in the main text we choose an adaptive truncation threshold and maximum bond dimension such that the discarded weight is always $\lesssim 10^{-10}$ with $\chi \le 1200$ (in practice $\chi$ increases with the tilt angle $\theta$, the subsystem size $N_A$, and decreasing $\alpha$). With this parameter schedule, we perform multiple random sampling iterations for each
fixed parameter set $(N,\theta,\alpha,N_A)$ and estimate
$\overline{P_A^{(2)}}$, $\overline{P_{A,Q_A}^{(2)}}$, and the corresponding entanglement asymmetry. The numerical results for Eq.~\eqref{numerics} are obtained by averaging over $50$ independent sampling iterations. All calculations are carried out using the \texttt{ITensor} library~\cite{itensor1,itensor2} and \texttt{TensorCircuit-NG}~\cite{Tensorcircuit}.

\section{Additional numerical results}\label{AppendixB}
In this section, we provide additional numerical results to illustrate the entanglement asymmetry dynamics of the initial states TFS, TNS, and TDWS across a broader range of $\alpha$ values, as well as the fate of the QME.

\begin{figure*}[bt]
\hspace*{-0.8\textwidth}
\centering
\includegraphics[width=0.8\linewidth]{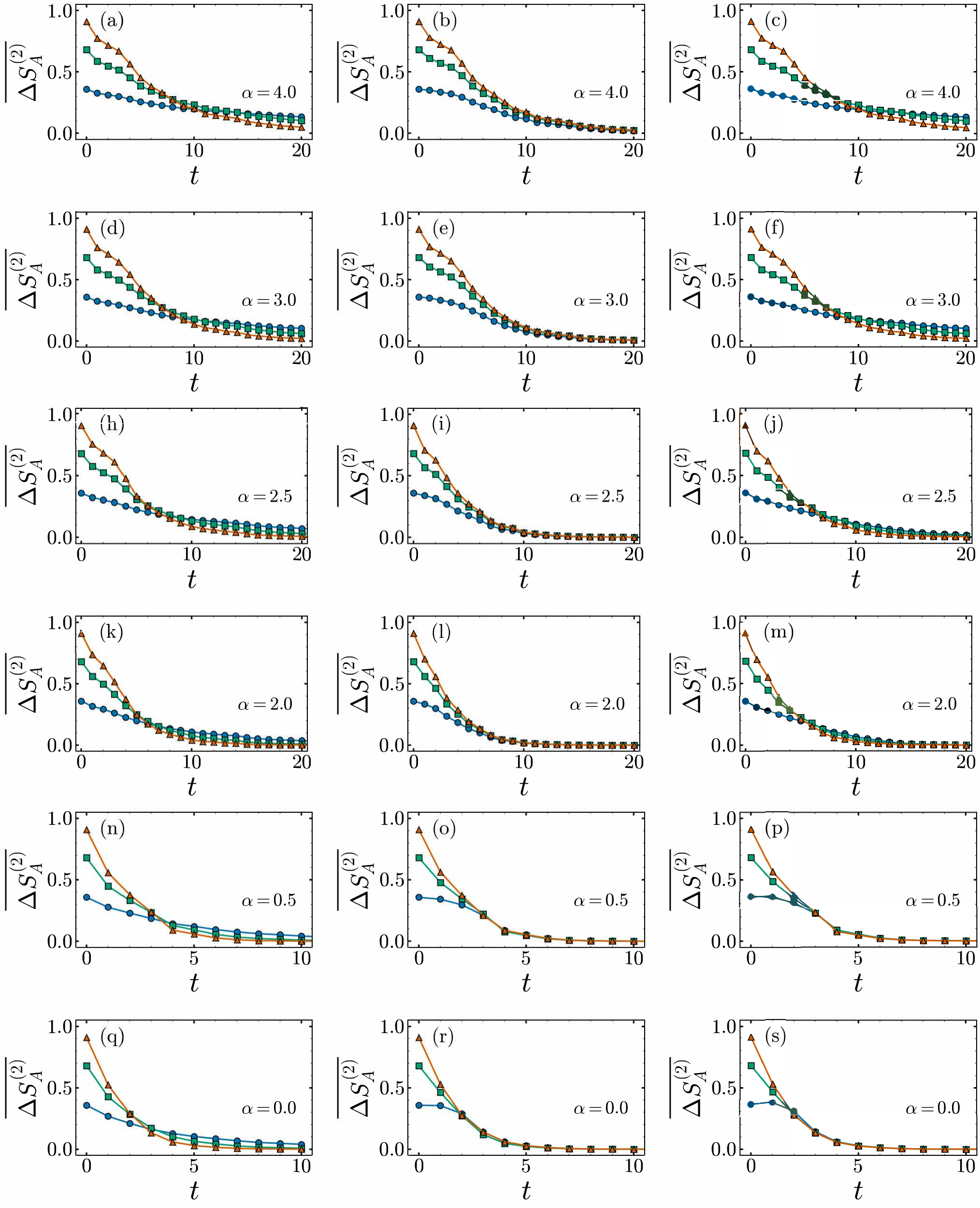} 
\caption{ The dynamics of entanglement asymmetry are simulated using replica tensor-network methods, TFS, TNS, and TDWS, for a system size of $N=48$ and subsystem size of $N_A=2$ with $50$ ensembles averaging. Specifically: TFS is shown in panels (a), (d), (h), (k), (n), and (q) for $\alpha = 4, 3, 2.5, 2.0, 0.5, 0$, respectively. TNS is shown in panels (b), (e), (i), (l), (o), and (r) for the same $\alpha$ values. TDWS is shown in panels (c), (f), (j), (m), (p), and (s) for the same $\alpha$ values.}\label{fig:EA_SM}
\end{figure*}

In the main text, we have presented numerical results with $\alpha = 5$ and $\alpha = 1.5$. Here, we show the entanglement asymmetry evolution for:
\begin{itemize}
    \item TFS [Fig.~\ref{fig:EA_SM} (a), (d), (h), (k), (n), (q)] at $\alpha = 4,\ 3,\ 2.5,\ 2.0,\ 0.5,\ 0$;
    \item TNS [Fig.~\ref{fig:EA_SM} (b), (e), (i), (l), (o), (r)] at $\alpha = 4,\ 3,\ 2.5,\ 2.0,\ 0.5,\ 0$;
    \item TDWS [Fig.~\ref{fig:EA_SM} (c), (f), (j), (m), (p), (s)] at $\alpha = 4,\ 3,\ 2.5,\ 2.0,\ 0.5,\ 0$,
\end{itemize}
with system size fixed at $N=48$ and subsystem size $N_A=2$.

As shown in Fig.~\ref{fig:EA_SM}, the TFS (first column) exhibits an $\alpha$-independent QME; the TNS (second column) shows no QME for any $\alpha$; and the TDWS (third column) displays an $\alpha$-dependent QME---specifically, the QME is absent for long-range interactions (efficiently small $\alpha$) but present for short-range interactions (efficiently large $\alpha$).

\section{Benchmark for replica tensor-network and exact diagonalization\label{AppendixC}}
\begin{figure*}[bt]
\hspace*{-0.6\textwidth}
\centering
\includegraphics[width=0.6\linewidth]{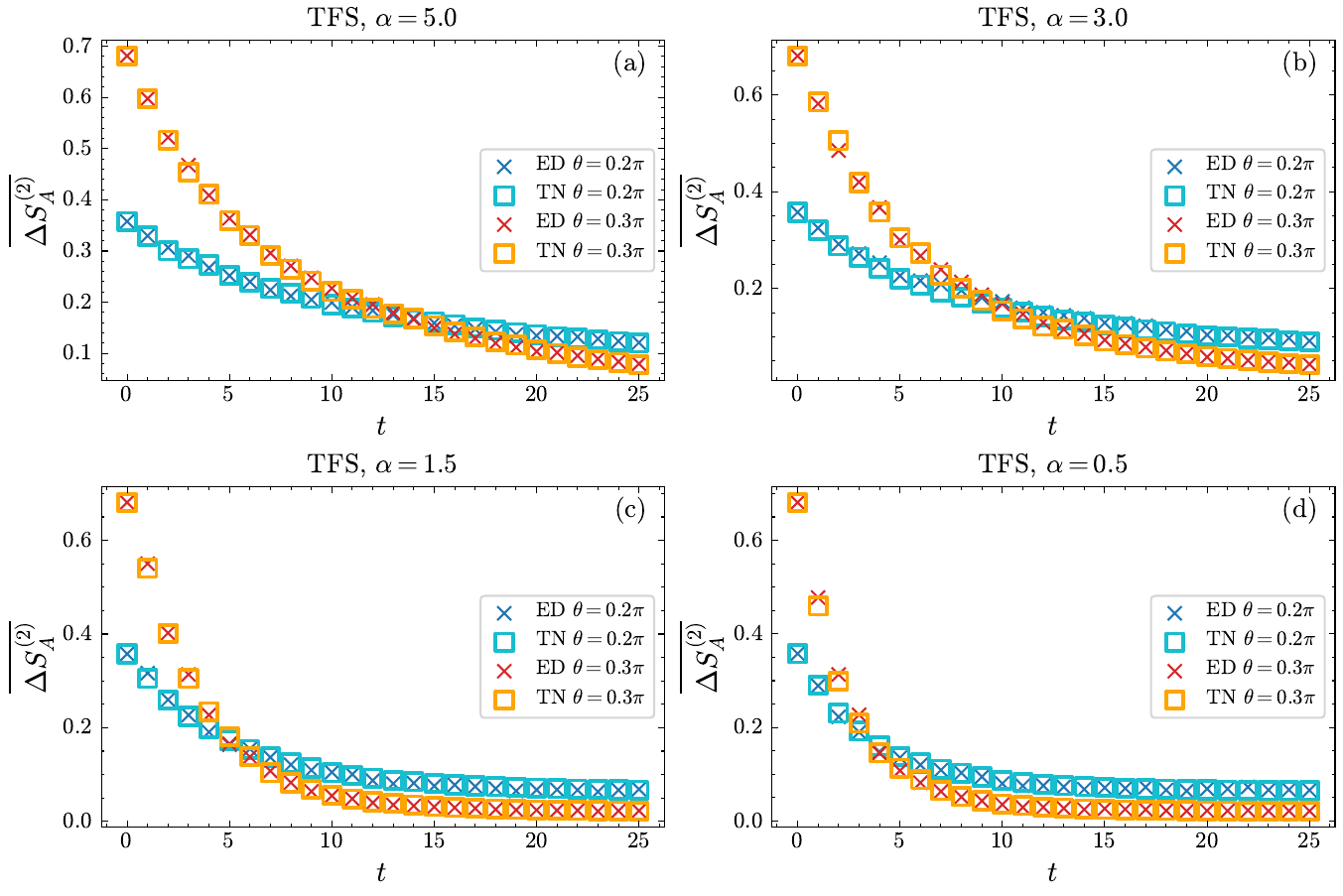} 
\caption{Time evolution of the annealed R\'enyi-2 entanglement asymmetry 
$\Delta S_A^{(2)}(t)$ for the tilted ferromagnetic state (TFS) at different power-law exponents 
$\alpha=5.0,\,3.0,\,1.5,\,0.5$ in panels (a)-(d), respectively. 
Results are shown for chain length $N=12$ and subsystem size $N_A=2$. Crosses (ED) denote exact diagonalization data, while open squares (TN) denote results from 
replica tensor-network simulations. Two tilt angles, $\theta=0.2\pi$ and $\theta=0.3\pi$, 
are compared, showing consistent relaxation behavior between the two numerical methods.}
\label{fig:benchmark}
\end{figure*}
To validate the replica tensor-network (TN) approach introduced above, we benchmark it against exact diagonalization (ED) for small system sizes. In Fig.~\ref{fig:benchmark}, we compare the time evolution of the annealed R\'enyi-2 entanglement asymmetry $\overline{\Delta S_A^{(2)}}(t)$ for the TFS at different power-law exponents $\alpha=5.0,\,3.0,\,1.5$, and $0.5$. The data correspond to a chain of length $N=12$ and a subsystem size $N_A=2$. We choose two tilt angles, $\theta=0.2\pi$ and $\theta=0.3\pi$ to test the dependence on the initial state. For all values of $\alpha$, the TN results (open squares) are in excellent agreement with the ED data (crosses) throughout the entire time window, accurately capturing both the short-time decay and the asymptotic saturation of $\overline{\Delta S_A^{(2)}}(t)$. Small discrepancies at very early times originate from finite sampling in the stochastic TN averaging, but they quickly vanish as the system approaches its steady state. These benchmarks confirm that the replica TN method faithfully reproduces the full quantum dynamics of $\overline{\Delta S_A^{(2)}}(t)$ in the accessible regime, and can therefore be reliably applied to much larger systems beyond the reach of ED.

\end{document}